\documentclass[twocolumn,amsmath,amssymb,pre,showpacs]{revtex4}

\usepackage{graphicx}
\usepackage{dcolumn}
\usepackage{bm}
\usepackage{amssymb}
\usepackage{amsmath}
\usepackage{amsthm,amsfonts}
\begin{document}
\bibliographystyle{aip}

\title{Beyond the Lindblad Master Equation: Heat, Work and Energy Currents in Boundary Driven Spin Chains}

\author{Lu\'is H. Reis, Saulo H. S. Silva, and Emmanuel Pereira}
 \email{emmanuel@fisica.ufmg.br}
\affiliation{Departamento de F\'{\i}sica--Instituto de Ci\^encias Exatas, Universidade Federal de Minas Gerais, CP 702,
30.161-970 Belo Horizonte MG, Brazil}

\begin{abstract}
We consider the accurate investigation of the energy current and its components, heat and work, in some boundary driven quantum spin systems.
The expressions for the currents, as well as the associated Lindblad master
equation, are obtained via a repeated interaction scheme. We consider small
systems in order to analytically compute the steady distribution to study
the current in the steady state. Asymmetrical $XXZ$ and quantum Ising models
are detailed analyzed. For the $XXZ$ chain we present cases in which different
compositions of heat and work currents, obtained via the repeated interaction
protocol, lead to the same energy current, which may be obtained via the Lindblad master equation. For the quantum Ising chain, we describe a case
of zero energy current and novanishing heat and work currents. Our findings
make clear that to talk about heat in these boundary driven spin quantum
systems we must go beyond an investigation involving only the Lindblad master
equation.
\end{abstract}


\def \Z {\mathbb{Z}}
\def \R {\mathbb{R}}
\def \La {\Lambda}
\def \la {\lambda}
\def \ck {l}
\def \F {\mathcal{F}}
\def \M {\mathcal{M}}
\newcommand {\md} [1] {\mid\!#1\!\mid}
\newcommand {\be} {\begin{equation}}
\newcommand {\ee} {\end{equation}}
\newcommand {\ben} {\begin{equation*}}
\newcommand {\een} {\end{equation*}}
\newcommand {\bg} {\begin{gather}}
\newcommand {\eg} {\end{gather}}
\newcommand {\ba} {\begin{align}}
\newcommand {\ea} {\end{align}}
\newcommand {\tit} [1] {``#1''}

%%%%%%%%%%%%%%%%%%%%%%%%%%%%%%%%%%%

\maketitle

\let\a=\alpha \let\b=\beta \let\d=\delta \let\e=\varepsilon
\let\f=\varphi \let\g=\gamma \let\h=\eta    \let\k=\kappa \let\l=\lambda
\let\m=\mu \let\n=\nu \let\o=\omega    \let\p=\pi \let\ph=\varphi
\let\r=\rho \let\s=\sigma \let\t=\tau \let\th=\vartheta
\let\y=\upsilon \let\x=\xi \let\z=\zeta
\let\D=\Delta \let\G=\Gamma \let\L=\Lambda \let\Th=\Theta
\let\P=\Pi \let\Ps=\Psi \let\Si=\Sigma \let\X=\Xi
\let\Y=\Upsilon

\section{Introduction}

The investigation of the energy transport laws  is a fundamental issue of nonequilibrium statistical physics \cite{BLiRMP, LLP, Dhar}. In particular, the study of the energy
transport properties at quantum scale is a problem that interests to experimental and theoretical researchers and that is receiving increasing attention nowadays \cite{BP, GZ}. Such study  is recurrent, for example in quantum spin chains, and its interest is enhanced by several different problems and motivations: the emerging
field of quantum thermodynamics, the advance of lithography and the possibility to manipulate small quantum systems, the properties of cold atoms and related phenomena, the possibility
of different regimes of transport in condensed matter, the possibility of rectifiers, etc.

Open quantum spin chains, such as the $XXZ$ $1D$ systems, are the archetypal models of open quantum systems \cite{xxz1, xxz2, xxz3}, that associates  to different problems in  nonequilibrium statistical physics, optics, quantum information, etc., they are
exhaustively investigated. In particular their boundary driven versions, i.e., systems with target spin polarizations at the boundaries, are recurrently studied. In opposition to the version in which the system is
weakly coupled to the baths, these boundary driven systems involve a process which includes work, not just only heat \cite{FBarra, Pereira2018, GL-NJP}. The weakly coupled version, otherwise, involves a work-free process. 

The
split of the energy current into heat and power is ignored in many articles in the literature, which may lead to incorrect conclusions \cite{EPL-Levy}.
 In fact, for thermodynamical consistency, such a decomposition of the energy
 is mandatory. In Ref.\cite{EPL-Levy}, that is entitled ``The local approach
 to quantum transport may violate the second law of thermodynamics'', the 
 authors treat transport in a system of two coupled harmonic oscillators. When considering the energy as only heat, they find this thermodynamical problem. However, with the decomposition of energy into heat and work, as detailed
 performed in Ref.\cite{GL-NJP}, it is shown that there is no inconsistency in
 the oscillator system, which may operate as a refrigerator.
 
 In a previous paper \cite{Pereira2018}, we stressed the
distinction between heat and work (power) in the energy current of the boundary driven $XXZ$ chain: by using the repeated interaction (RI)
protocol \cite{RI, EspoPRX, LOK}, we derived algebraic expressions for the heat and work currents
that showed that the final energy current given by the sum of heat and power was in agreement with the energy current expression obtained via the usual continuity equation.

In the present paper we focus on the details of the energy current components, i.e., by computing exactly  the density matrix of some small systems, namely, $XXZ$ and quantum Ising models, we perform analytical investigations and describe in details
the values and behavior of the heat and power currents in different situations (different parameters for the interacting systems, for the external baths, etc.), a problem still to solve.

It is pertinent, as already said, to emphasize the importance of the distinction of heat and work (power) in the energy current of these boundary driven  spin systems. The final dynamics given by the usual Lindblad master equation (LME) allows
us to describe the total energy current only, where the action of the baths is described in terms of certain dissipators which involve the driving strength related to the bath spin polarization at the boundaries. However,
when we start the analysis considering the whole repeated interaction process, which leads later to the final LME and also to the equations for heat and work, we observe that different processes (e.g., with different compositions
of heat and work) can lead to the same LME and to the same energy current, as
we show in the present paper. In other words, more information is obtained  with the consideration of the repeated interaction protocol.

Here, in the present article, we investigate asymetrical $XXZ$ and quantum Ising models, i.e., systems with different (asymmetrical) intersite interactions. We perform analytical studies to compute the steady state density of such models. For the $XXZ$ chains, we show the existence of different decompositions into heat and work leading to the same energy current. For the quantum Ising chain, we show a case of zero energy flow but
nonvanishing heat and work currents. Our results clarify that to talk about heat in these recurrently used quantum spin boundary driven systems we
must go beyond the associated LME.

We still want to say that versions of these quantum spin models can be experimentally realized. It is possible to engineer quantum
$XXZ$ Hamiltonian with different values for the inner parameters \cite{Endres, Barredo}, and there are recent experimental works with Rydberg
atoms in optical traps \cite{Duan, Ng} related to $XXZ$ models.

The rest of the paper is organized as follows. In section 2, we present the formalism: the repeated interaction protocol; the LME for target $\sigma^{z}$ polarization at the boundaries; the associated formulas for the currents. In section 3, we describe the results for the currents in the cases of the $XXZ$ model and the quantum Ising model. Section 4 is devoted for final remarks, and
the appendix to some technical points.

\section{The Repeated Interactions Protocol}

The RI protocol is our basis for the dynamical investigations, it is the framework to be used in the derivation of the currents for heat and work, as well as of the LME associated to the problem.

As described in Ref.\cite{Landi2}, the inspiration for the RI protocol comes from the Boltzmann's "Stosszahlansatz": at any given time interval, the 
system interacts with only a small fraction of the bath. A clear example is
given by the Brownian motion, in which a particle interacts with some few
water molecules in a given time, and for a very short time. After this time,
the molecules go away and do not return. The bath is large, and so, the next
molecule to interact is completely uncorrelated from the previous one. And
the process repeats again. 

We describe the RI scheme in details. We start from a system with time-independent Hamiltonian $H_{S}$ coupled to two baths, the left and the right one, with Hamiltonian $H_{L}$ and $H_{R}$. We divide the time scale into intervals of size
$\tau$, i.e., intervals $t \in [(n-1)\tau, n\tau)$. At time zero, and also at the beginning of each time interval, we make the assumption of system and baths decoupled, i.e., for the total density matrix we take
$$
\rho_{\rm total}(0) = \rho_{S} \otimes \rho_{E}~,
$$
where $E$ (environment) denotes the left and right baths. Then, with the baths coupled to the system, we allow the whole set to evolve up to a time $\tau$. After such an evolution, we take the partial trace over the baths
to obtain $\rho_{S}(\tau)$. Then we couple the resulting system to a new copy of left and right baths. The whole set is allowed to evolve from time $\tau$ to time $2\tau$. Once again we take the partial trace over the baths
and repeat, indefinitely, the same process.

Let us write the expressions. First, we take a collection of Hamiltonian for the baths, i.e., we write $H_{r} = \sum H^{n}_{r}$, where $r$ is $R$ or $L$, and $n \in [1,2,3,\ldots]$. Each $H^{n}_{r}$ interacts with the system
for times $t \in [(n-1)\tau, n\tau)$. Similarly, we write the interaction baths-system as $V(t) = \sum_{n}V^{n}$, with $V^{n} = V^{n}_{L} +  V^{n}_{R}$, again $n \in [1,2,3,\ldots]$. The density matrices for the baths are denoted by
$\rho_{E} = \otimes_{n} \rho_{n}$, where, at the beginning of each time interval we assume a Boltzmann-Gibbs distribution for the baths
\begin{eqnarray*}
 \rho_{n} &=& \omega_{\beta_{L}}(H^{n}_{L})\otimes  \omega_{\beta_{R}}(H^{n}_{R})~, \\
  \omega_{\beta_{r}} &=& e^{-\beta H_{r}}/(Tr_{r} e^{-\beta H_{r}})~.
 \end{eqnarray*}
Thus, according to the RI protocol, the density matrix evolves following the map
\begin{equation} \label{map}
\rho_{S}(n\tau) = Tr_{n}\{ U_{n}[\rho_{S}((n-1)\tau)\otimes\rho_{n}]U_{n}^{\dagger}\} ~,
\end{equation}
where $Tr_{n}$ means the trace over the copy $n$ of the baths, and
$$
U^{n} = \exp[-i\tau H_{tot}] = \exp[ -i\tau( H_{S} + H^{n}_{L} + H^{n}_{R} + V^{n})]~.
$$
We take $\hbar = 1$.

The LME may be obtained by expanding the map equations up to first order in $\tau$ (and properly redefining $V_{L}$ and $V_{R}$ as proportional to $1/\sqrt{\tau}$, see Appendix). One obtains
\begin{equation}
\frac{d\rho_{S}}{dt} = -i[H_{S}, \rho_{S}] + \mathcal{D}_{L}(\rho_{S}) + \mathcal{D}_{R}(\rho_{S})~,
\end{equation}
where $\mathcal{D}_{L}$ (and $\mathcal{D}_{R}$) is related to $-Tr_{n}[V_{L},[V_{L},\rho_{S}]]$, with $[\cdot, \cdot]$ meaning the commutator. For a spin bath
$$
V_{L} = \sqrt{\frac{\gamma_{L}}{\tau}} \left(\sigma_{L}^{x}\sigma_{1}^{x} + \sigma_{L}^{y}\sigma_{1}^{y}\right)~,
$$
where $\gamma_{L}$ is the coupling strength to the bath, 
similarly for $V_{R}$ (with $\sigma_{R}, \sigma_{N}$ instead of $\sigma_{L}, \sigma_{1}$), and $H_{L} = h_{L}\sigma_{L}^{z}/2$ (similarly for $H_{R}$), we obtain the following LME
\begin{eqnarray}\label{LME}
\dot{\rho_{S}} &=& -i[H_{S}, \rho_{S}] + \mathcal{D}_{L}(\rho_{S}) + \mathcal{D}_{R}(\rho_{S})~, \nonumber\\
\mathcal{D}_{L,R}(\rho_{S}) &=& \sum_{k=\pm} L_{k}\rho_{S}L_{k}^{\dagger} - \frac{1}{2}\left\{L^{\dagger}_{k}L_{k},\rho_{S}\right\}~,
\end{eqnarray}
where, in $\mathcal{D}_{L}$ we have
\begin{equation}
L_{\pm} = \sqrt{2\gamma_{L}(1 \pm f_{L})} \sigma_{1}^{\pm}~.
\end{equation}
And similarly in $\mathcal{D_{R}}$, which involves $\gamma_{R}, f_{R}, \sigma_{N}^{\pm}$. In the equation above,  $f_{L} = \left< \sigma_{L}^{z}\right>$ and
$f_{R} = \left< \sigma_{R}^{z}\right>$ are the bath spin polarizations at the edges; $\sigma_{j}^{\pm} \equiv (\sigma_{j}^{x} \pm i\sigma_{j}^{y})/2$ are the spin creation and annihilation operators. We will take, in what
follows, $\gamma_{L} = \gamma_{R} = \gamma$. In terms of $\sigma_{j}^{\pm}$, the dissipator in the LME becomes
\begin{widetext}
\begin{eqnarray}
\mathcal{D}(\rho_{S}) &=& \gamma\left\{(1+f_{L})\left[2\sigma_{1}^{+}\rho_{S}\sigma_{1}^{-} - \left( \sigma_{1}^{-}\sigma_{1}^{+}\rho_{S}  + \rho_{S} \sigma_{1}^{-}\sigma_{1}^{+}\right)\right]\right.
                      + (1-f_{L})\left[2\sigma_{1}^{-}\rho_{S}\sigma_{1}^{+} - \left( \sigma_{1}^{+}\sigma_{1}^{-}\rho_{S}  + \rho_{S} \sigma_{1}^{+}\sigma_{1}^{-}\right)\right]  \nonumber\\
 &&  + (1+f_{R})\left[2\sigma_{N}^{+}\rho_{S}\sigma_{N}^{-} - \left( \sigma_{N}^{-}\sigma_{N}^{+}\rho_{S}  + \rho_{S} \sigma_{N}^{-}\sigma_{N}^{+}\right)\right]
                    \left.  + (1-f_{R})\left[2\sigma_{N}^{-}\rho_{S}\sigma_{N}^{+} - \left( \sigma_{N}^{+}\sigma_{N}^{-}\rho_{S}  + \rho_{S} \sigma_{N}^{+}\sigma_{N}^{-}\right)\right] \right\}~,
\end{eqnarray}
\end{widetext}
where $\mathcal{D} = \mathcal{D}_{L} + \mathcal{D}_{R}$.

To describe the expressions for the currents of heat and work we follow F. Barra \cite{FBarra}; see also ref.\cite{Pereira2018}. We briefly resume the procedure here.

For a system coupled with an environment, the internal energy is defined as
\begin{equation}
E(t) = Tr(\rho_{tot}(t)[H_{S}(t) + V(t)]) ~,
\end{equation}
where $Tr$ denotes the full trace. According to the first law of thermodynamics, the change in $E$ is involved with heat and work
$$
\Delta E(t) = W(t) + Q(t)~,
$$
where the heat $Q(t) = \sum_{r} Q_{r}(t)$, $r= R, L$, that flows to the system in the time interval $[0,t]$ is defined as
\begin{equation}
Q_{r}(t) = Tr( H_{r}\rho_{tot}(0) - H_{r}\rho_{tot}(t))~,
\end{equation}
which means minus the change in the energy of bath $r$. Consequently, the work performed on the system in the interval $[0,t]$ is given by
\begin{equation}
W(t) = Tr(\rho_{tot}(t)H_{tot}(t) - \rho_{tot}(0)H_{tot}(0))~.
\end{equation}
The definition of work as the change in the total energy, as well as the definition of heat, is intuitive. Recall that in a classical mechanical system, the work performed by the
system is $W = F\Delta x = -(\Delta H/\Delta x) \Delta x$. 

To carry out the analysis within the repeated interactions protocol, we take time intervals. For $t \in [(n-1)\tau, n\tau)$ (we use the index $n$ in some operators to refer to such interval), we have
$$
\Delta Q_{r} = Tr(H^{n}_{r}(\rho_{n} - \rho_{n}'))~,
$$
where
\begin{eqnarray*}
\rho_{n}' &=& Tr_{S}(U_{n}\rho_{S}((n-1)\tau)\otimes\rho_{n} U^{\dagger}_{n})~,\\
U_{n} &=& e^{i\tau(H_{S} + H^{n}_{L} + H^{n}_{R} + V^{n})}~.
\end{eqnarray*}
Then, writing $V^{n}= v^{n}/\sqrt{\tau}$, expanding $U^{n}$ in powers of $\tau$, after some manipulations (see \cite{FBarra}) we obtain
\begin{widetext}
\begin{equation}
\dot{Q}_{r} = \lim_{\tau\rightarrow 0} \frac{\Delta Q_{r}}{\tau} =  - Tr\left( \left(v_{r}H_{r}v_{r} - \frac{1}{2}\{v_{r}^{2},H_{r}\}\right)\rho_{S}(t)\otimes\omega_{\beta_{r}}\right) ~,
\end{equation}
\end{widetext}
where $\{\cdot, \cdot\}$ above is the anticommutator, and we omitted the superscript in $v$. For our specific $XXZ$ model, we have (for $r =L$)
\begin{equation}
\dot{Q}_{L} = 2\gamma_{L}h_{L} \left[ f_{L} - Tr_{S}\left(\sigma_{1}^{z}\rho_{S}(t)\right)\right]~,
\end{equation}
where
\begin{equation*}
f_{L} \equiv Tr_{L}\left(\sigma_{L}^{z}\omega_{\beta_{L}}\right) = -\tanh\left(\beta_{L}\frac{h_{L}}{2}\right) ~.
\end{equation*}

For the expression for the work variation, we make a shift in the time interval and analyze the work between the times $n\tau - \epsilon$ and $n\tau + \epsilon$ (when a bath is replaced by a new one and we exchange the potential $V^{n}$ to
$V^{n+1}$). Hence, $\Delta W = \Delta W_{L} + \Delta W_{R}$, with $\Delta W_{L} = Tr([ V^{n+1}_{L} - V^{n}_{L}]\rho_{tot})$.  Repeating an analysis similar to that performed for $\dot{Q}_{L}$ (see \cite{FBarra}), we obtain
 \begin{widetext}
\begin{equation}
\dot{W}_{r} = \lim_{\tau\rightarrow 0} \frac{\Delta W_{r}}{\tau} = Tr\left( \left(v_{r}(H_{S} + H_{r})v_{r} - \frac{1}{2}\{v_{r}^{2}, H_{S}+H_{r}\}\right)
\rho_{S}(t)\otimes\omega_{\beta_{r}}\right) ~.
\end{equation}
For the specific $XXZ$ model, we have
\begin{eqnarray}
\dot{W}_{L} &=& 2h_{1}\gamma_{L}\left[f_{L} - Tr_{S}(\sigma_{1}^{z}\rho_{S}(t))\right] -
2h_{L}\gamma_{L}\left[f_{L} - Tr_{S}(\sigma_{1}^{z}\rho_{S}(t))\right] \nonumber\\
& & - 2\gamma_{L} Tr_{S}\left( [\alpha(\sigma_{1}^{x}\sigma_{2}^{x} + \sigma_{1}^{y}\sigma_{2}^{y}) +
\Delta_{1,2}\sigma_{1}^{z}\sigma_{2}^{z}]\rho_{S}(t)\right) \nonumber \\
& & - 2\gamma_{L}\Delta_{1,2}Tr_{S}\left(\sigma_{1}^{z}\sigma_{2}^{z}\rho_{S}(t)\right) + 4\gamma_{L}\Delta_{1,2}f_{L}Tr_{S}\left(\sigma_{2}^{z}\rho_{S}(t)\right) ~.
\end{eqnarray}
\end{widetext}

By adding the terms $\dot{W}_{L}$ and $\dot{Q}_{L}$ and the similar ones for the right end, we obtain the expression for the energy rate
$$
\dot{E} = \dot{W}_{L} + \dot{Q}_{L} + \dot{W}_{R} + \dot{Q}_{R}~.
$$

Now we recall an interesting result. When we study the energy current directly from the LME we take the continuity equation (meaning the change in the energy bond linking sites $i$ and $i+1$ equals the difference between
the energy that comes to the site $i$ from the left sites and the energy that leaves the site $i+1$ to the right sites)
$$
\frac{\langle \varepsilon_{i,i+1}\rangle}{dt} = - \left( \langle F_{i+1}\rangle - \langle F_{i}\rangle\right)~,
$$
where $\varepsilon_{i,i+1}$ comes from the Hamiltonian splitting, e.g., for the $XXZ$ model
\begin{eqnarray}
H_{S} &=&  \sum_{i=1}^{N-1} \varepsilon_{i,i+1} = \sum_{i=1}^{N-1} h_{i,i+1} + b_{i,i+1} ~,  \\
h_{i,i+1} &=& \alpha \left( \sigma_{i}^{x} \sigma_{i+1}^{x} + \sigma_{i}^{y}\sigma_{i+1}^{y} \right) + \Delta_{i,i+1} \sigma_{i}^{z} \sigma_{i+1}^{z} ~, \nonumber\\
b_{i,i+1} &=& \frac{1}{2} \left[ \frac{h_{i}}{2}\sigma_{i}^{z}(1+\delta_{i,1}) + \frac{h_{i+1}}{2}\sigma_{i+1}^{z}(1+\delta_{i+1,N}) \right]~.\nonumber
\end{eqnarray}
In the steady state, $d\langle\cdot\rangle/dt = 0$, and the same current that arrives at one bond leaves it.
As shown in ref.\cite{Pereira2018}, the energy current that comes into the system is
$$
\langle F_{1} \rangle = \dot{W}_{L} + \dot{Q}_{L}~,
$$
that is, the energy current obtained directly from the LME and the continuity equation is the same that the one obtained via the split into heat and work in the repeated interaction protocol. However, from the LME we
cannot split the energy current. Moreover, as we show ahead, the repeated interaction protocol allows different processes (different values for work and heat) giving the same energy currents, in other words, the
same LME.

\section{Results}

{\it The asymmetrical $XXZ$ chain.} First we investigate asymmetrical $XXZ$ chains with $\sigma^{z}$ target polarization at the boundaries and without external magnetic field. In such a case, as computed in ref.\cite{SPL} and
explained in terms of symmetries of the LME in ref.\cite{Prapid2017}, we observe an interesting behavior of the energy current: the one-way street phenomenon. Precisely, as we invert the baths, i.e. the driving strengths $f$ and
$-f$ at the boundaries, the energy current does not change: it keeps its magnitude and also its direction (the current does not invert with the reservoirs inversion). As said, a physical explanation for the phenomenon
comes from symmetries in the LME: the expression for the energy current derived from the LME is an even function of $f$, and so, nothing changes as
with the baths inversion $f \leftrightarrow -f$. See Refs.\cite{Prapid2017, EPL2020}. 
The thermodynamic consistency of such phenomenon is
discussed in ref.\cite{Pereira2018} by considering the decomposition of the energy current into heat and work. In the present paper, by computing the steady state distribution of the $XXZ$ chain, we further on the analysis. Details about the computation of this steady density matrix are presented in the Appendix.

We first consider a small chain with three sites and Hamiltonian (in the absense of magnetic field)
\begin{equation}\label{Hamiltonian2}
H_{S} = \sum_{i=1}^{2} \alpha(\sigma_{i}^{x}\sigma_{i+1}^{x} + \sigma_{i}^{y}\sigma_{i+1}^{y}) + \Delta_{i,i+1}\sigma_{i}^{z}\sigma_{i+1}^{z}~,
\end{equation}
with  $\Delta_{1,2} = \Delta - \delta$, $\Delta_{2,3} = \Delta + \delta$.

In Fig.1 we plot the energy current $F$, the total work $W = W_{L} + W_{R}$, the heats $Q_{L}$ and $Q_{R}$ as functions of $\Delta$.  The one-way street phenomenon appears: the same figure follows if $f_{L}$ and $f_{R}$ are inverted ($f\rightarrow -f$). The inversion is obtained by keeping fixed $\beta_{L}$
and $\beta_{R}$, and inverting $h_{L}\rightarrow -h_{L}$ and $h_{R}\rightarrow -h_{R}$. No current changes its value.

\begin{figure}
\includegraphics[width=\columnwidth]{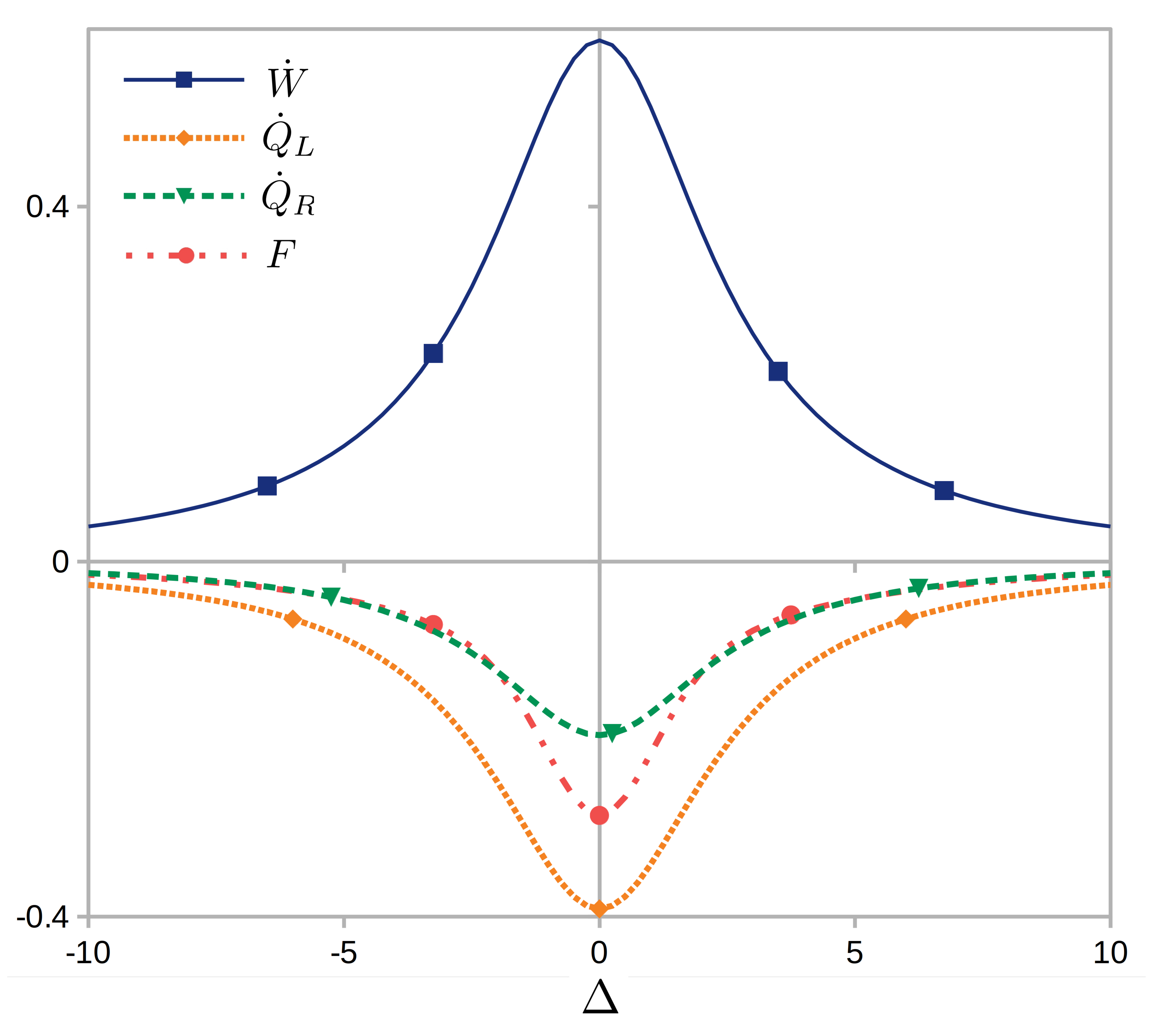}
\caption{(Color online)
Energy, left and right heat and work currents versus $\Delta$; we take $\gamma =\delta = 1$,
$h_L = 1$; $h_R = -0.5$; $\beta_R = 2$; $\alpha = 1$; the figure follows also for the case with inverted baths ($f\rightarrow -f$), in which we take $\gamma =\delta = 1$; $h_L = -1$; $h_R = 0,5$; $\beta_R = 2$; $\alpha = 1$.}
\label{Fig1}
\end{figure}

A comment is opportune here. In the expression of the LME (\ref{LME}), no 
parameter of the bath (i.e., temperature or magnetic field) is informed, only appears the driving strength $f$. The relation between $f$ and the inverse temperature $\beta$ and the field $h$ is given by the model of a bath spin polarization, i.e., by the averages of extra spins at the boundaries: $f_{L} = \langle \sigma_{L}^{z}\rangle = -\tanh(\beta_{L}h_{L}/2)$. Such a relation appears within the repeated interaction protocol, which will lead to the LME and also to
expressions for heat and work. Hence, one can see that it is possible to change $\beta_{L}$ and $h_{L}$ without changing $f_{L}$ (i.e., the product $\beta_{L}h_{L}$). A natural question here is that if we may have different
values for the currents of heat and work due to changes in $\beta_{L}$ and $h_{L}$ (or $\beta_{R}$ and $h_{R}$), but for fixed product $\beta_{L}h_{L}$, i.e., $f_{L}$. The answer is yes. In Fig.2 we plot, as in Fig.1, the currents
of energy, total work, left and right heat versus $\Delta$. But now, the inversion $f \rightarrow -f$ (see Fig.3) is performed by changing $h_{L} \rightarrow -h_{L}$ ($\beta_{L}$ is fixed), $h_{R} \rightarrow -h_{R}/10$ and
$\beta_{R} \rightarrow 10\beta_{R}$. Note that all currents change with the inversion, except the energy current which is the same.

We also observe changes in the currents, except energy, for other cases with the inversion $f \rightarrow -f$ and different combination of $h$ and $\beta$ (figures not plotted here).

Still concerning the one-way street phenomenon, we observe here its occurrence for the inversion of arbitrary $f_{L}$ and $f_{R}$, i.e., beyond the inversion between $f$ and $-f$ described in previous papers.
If we take $\alpha=\delta=1$, $h=\Delta=0$, and $\gamma_R = \gamma_L = 1$,  the energy current becomes
\begin{widetext}
\begin{equation}
F = -\frac{160 (e^{\beta_L h_L}-e^{\beta_R h_R})^2}{(e^{\beta_L h_L}+1)
(e^{\beta_R h_R}+1) (121 e^{\beta_L h_L}+117 e^{\beta_L h_L+\beta_R h_R}+121 e^{\beta_R h_R}+117)}~.
\end{equation}
\end{widetext}
From the expression above it is easy to note that if we arbitrarily change
 $f_L$ by $f_R$, that is, if we perform the changes $\beta_L \rightarrow \kappa_R \beta_R$, $h_L \rightarrow h_R/\kappa_R$, $\beta_R \rightarrow \kappa_L \beta_L$ e $h_R \rightarrow h_L/\kappa_L$, the energy current remains the same.
We note that
the energy current is the same, but the other currents can change depending on the choice for $\kappa_{L}$, $\kappa_{R}$, etc.

\begin{figure}
	\includegraphics[width=\columnwidth]{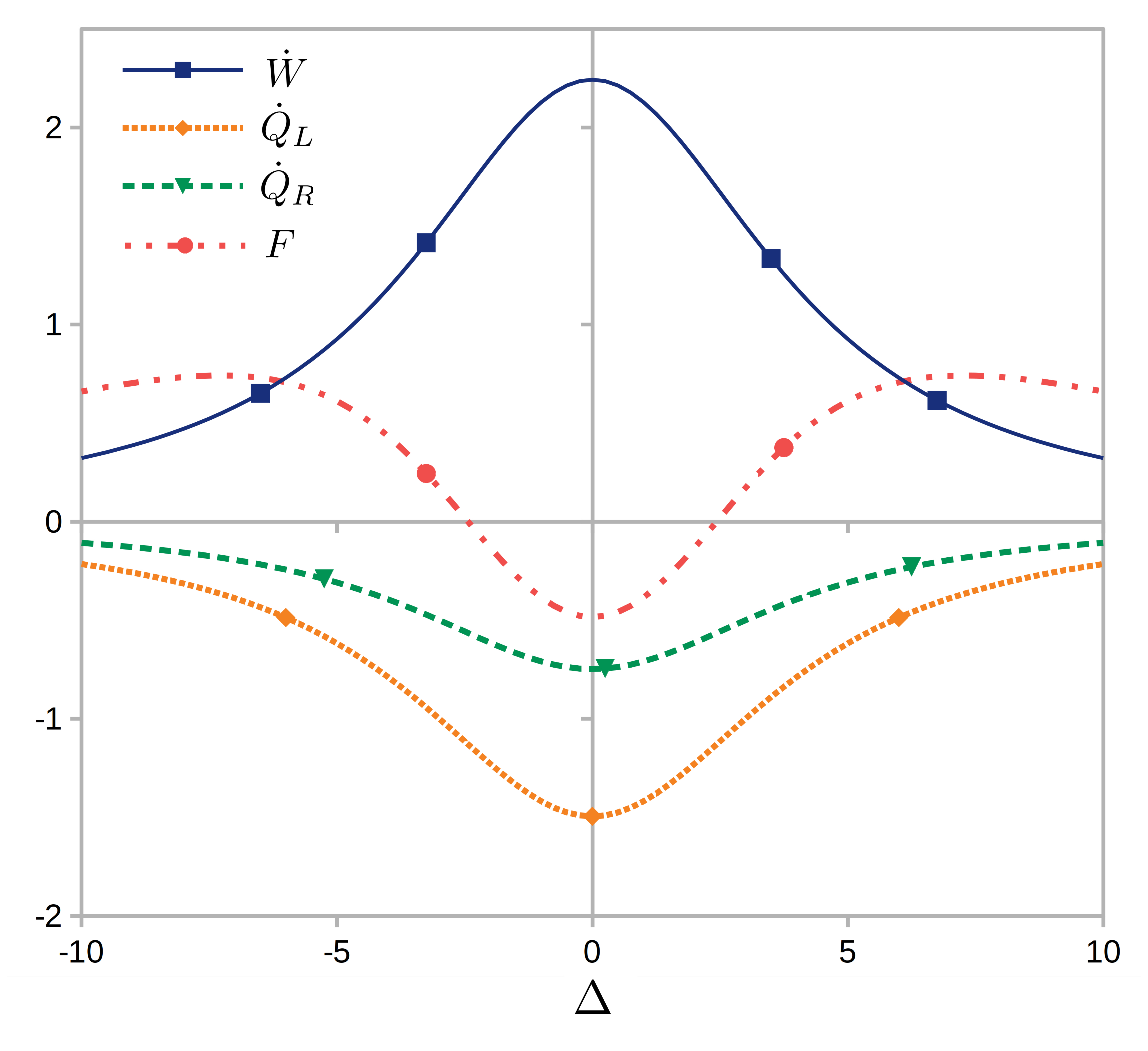}
	\caption{(Color online)
		Energy, left and right heat and work currents versus $\Delta$;  we have $\gamma =\delta = h_L = 1$, $h_R = -0.5$, $\beta_R = 10$, $\alpha = 2$.}
	\label{Fig2}
\end{figure}

\begin{figure}
	\includegraphics[width=\columnwidth]{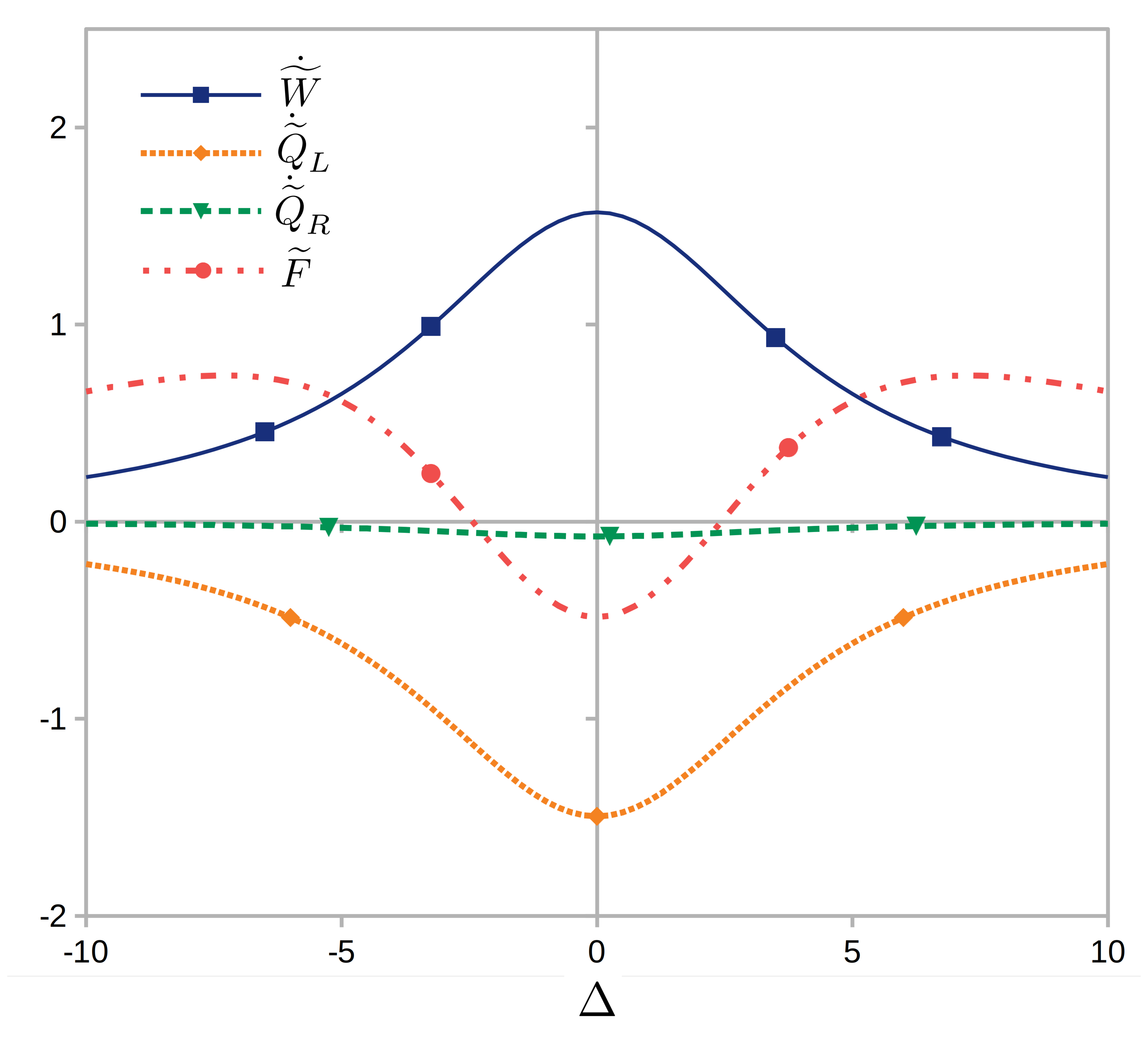}
	\caption{(Color online)
		Energy, left and right heat and work currents versus $\Delta$
		for the case of Fig.2 with inverted baths ($f\rightarrow -f$); we take $\gamma = \delta = 1$ , $h_L=-1$, $h_{R} =0.05$, $\beta_{R} = 100$, $\alpha = 2$.}
	\label{Fig3}
\end{figure}

One interesting point related to the repeated interaction protocol is that it leads to a steady state by considering a sequence of infinitesimal cycles, each one having an associated heat and work rate. In this sense, it is
profitable to search for conditions such that the system functions as a heat engine: as a refrigerator, thermal engine or heater. I.e., the refrigerator takes work to make heat flow from the cold bath to hot one:
$\dot{Q}_{1}>0$, $\dot{Q}_{2}<0$ and $\dot{W}>0$, where the index $1$ is for the colder end, $2$ for the hotter one. For the heater we have $\dot{Q}_{1}<0$, $\dot{Q}_{2}>0$ and $\dot{W}>0$; for the engine,
$\dot{Q}_{1}<0$, $\dot{Q}_{2}>0$ and $\dot{W}<0$.
In Fig.4 we show different regimes (as function of $h_{L}$) for the case of zero external magnetic field, where
we have the one-way street phenomenon, and in Fig.5 we plot different regimes for the system in the presence of an external, uniform magnetic field $h=1$,
i.e., with the addition in the previous Hamiltonian of $\sum_{i}h\sigma_{i}^{z}/2$.

\begin{figure}
\includegraphics[width=\columnwidth]{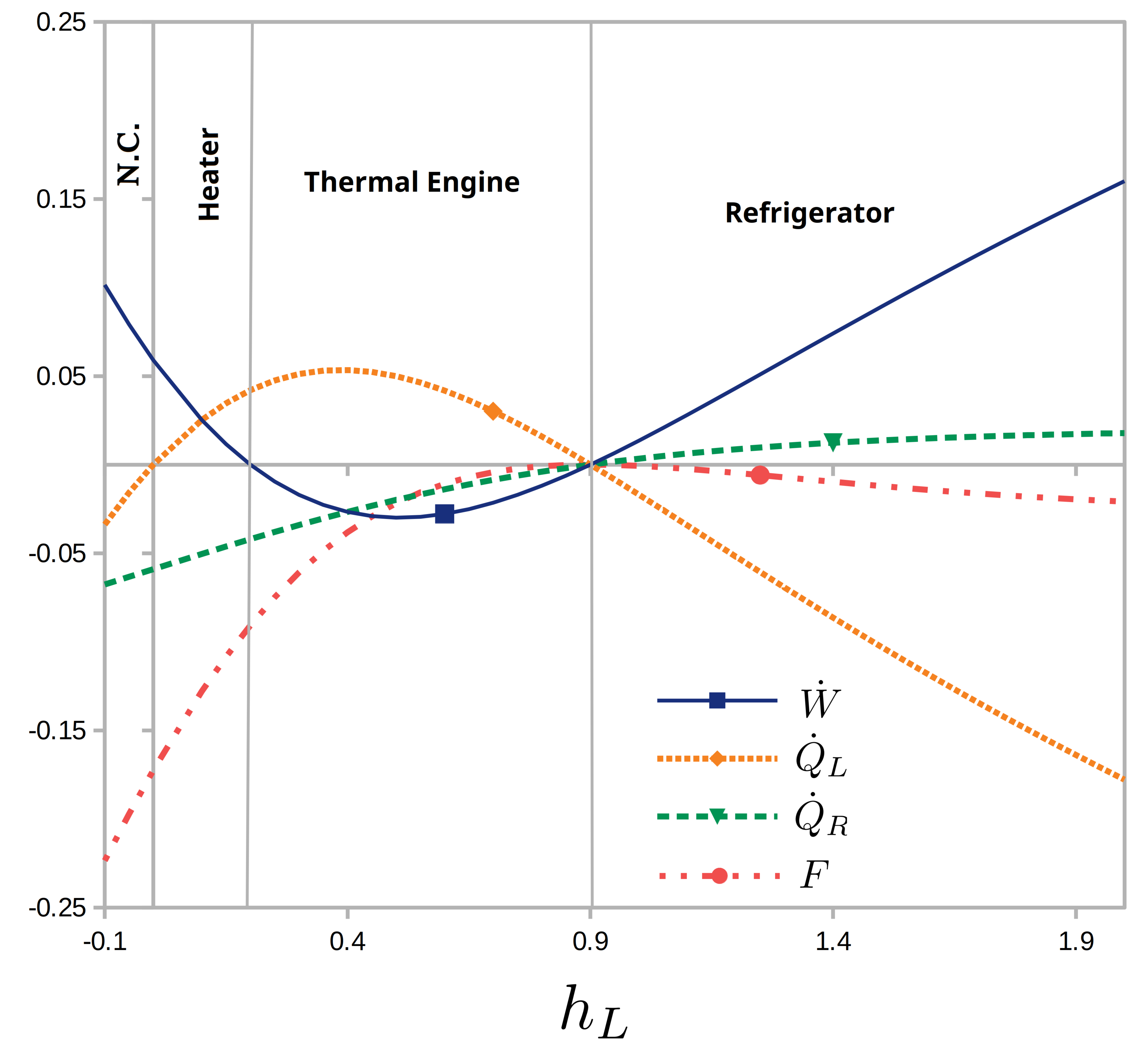}
\caption{(Color online) %(a-d) We plot the forward current $\cur_f$ (a), the reverse current $\cur_r$ (b), the rectification coefficient $\Rec$ (c), and the contrast $\Con$ (d) as functions of the anisotropy $\Delta_L/J_L$ for %different chain sizes $N$.
Different regimes as functions of $h_{L}$. We take $\alpha= \gamma= \delta= 1$, $h= \Delta= 0$, $h_{R}= 0.2, \beta_{R}=9, \beta_{L}=2$.}
\label{Fig4}
\end{figure}

\begin{figure}
\includegraphics[width=\columnwidth]{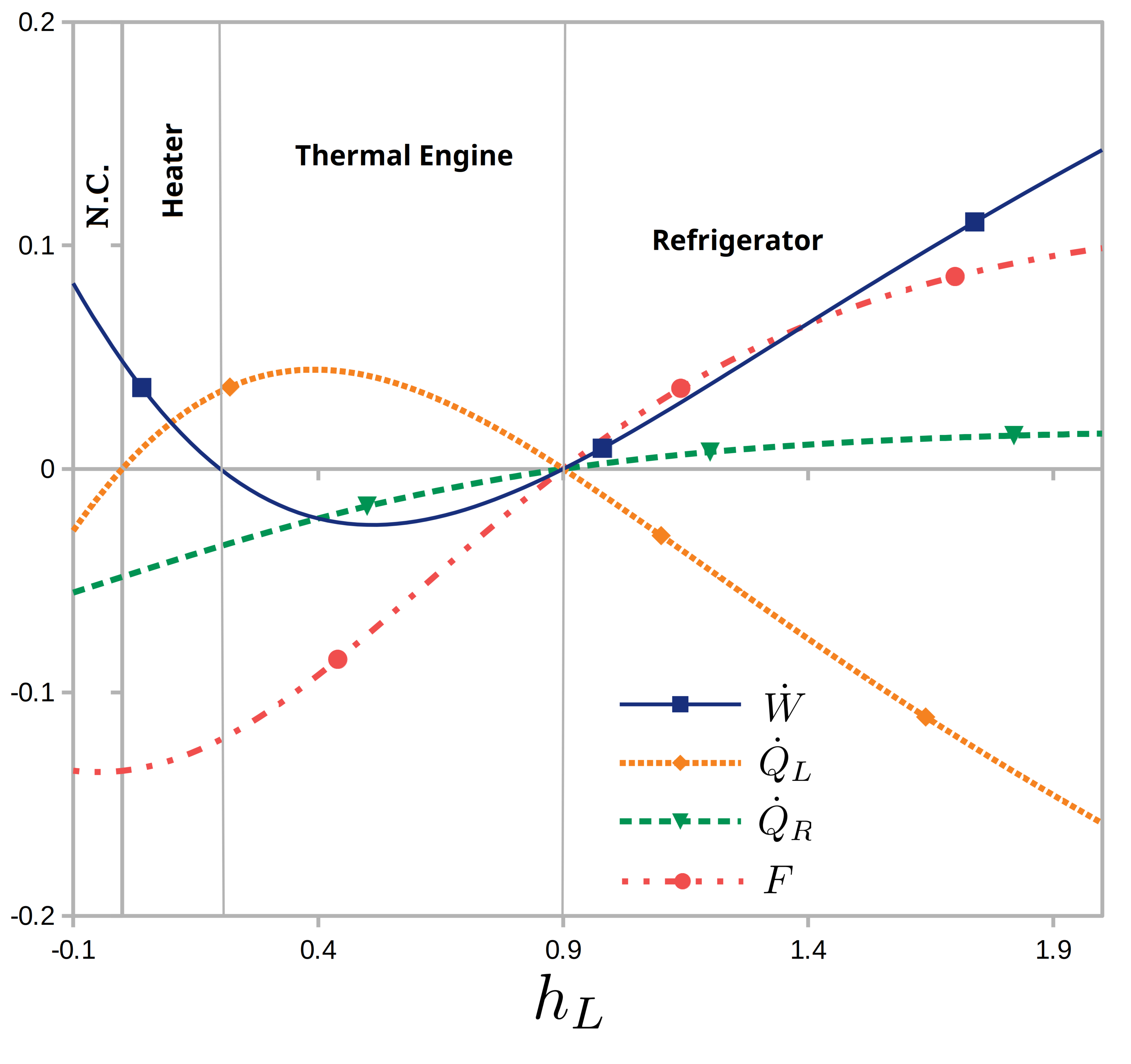}
\caption{(Color online) %(a-d) We plot the forward current $\cur_f$ (a), the reverse current $\cur_r$ (b), the rectification coefficient $\Rec$ (c), and the contrast $\Con$ (d) as functions of the anisotropy $\Delta_L/J_L$ for %different chain sizes $N$.
Different regimes as functions of $h_L$. We take $\alpha = \gamma =\delta = h = \Delta = 1$, $h_R = 0.2$, $\beta_R = 9$; $\beta_L = 2$.
%Different regimes as functions of $h_{L}$. We take $\alpha= \gamma= h= \delta= 1$, $\Delta= 0$, $h_{R}= 0.2, \beta_{R}=9, \beta_{L}=2$.
}
\label{Fig5}
\end{figure}

{\it The asymmetrical quantum Ising model.} Now we analyze the open quantum Ising model in the presence of a bath of bosons, and also the case of a spin bath. We want, again, to unveil the components of the energy current, namely,
heat and work.

We begin with the case involving a bath of bosons. For a system with $N=2$, we take the Hamiltonian
\begin{equation}
    H_{s}=\frac{h_{1}}{2}\sigma_{1}^{z}+\frac{h_{2}}{2}\sigma_{2}^{z}+\frac{\Delta_{1,2}}{2}\sigma_{1}^{z}\sigma_{2}^{z}~.
\end{equation}
For the baths we assume $H_{L(R)} = \omega_{L(R)}a^{\dagger}_{L(R)}a_{L(R)}$, where $a^{\dagger}$ and $a$ are the creation and annihilation boson operators; and for the interaction spin-bath we take $V_{L(R)} =
g_{L(R)}\sigma_{1(2)}^{x}(a_{L(R)}^{\dagger} + a_{L(R)})$, where $g$ is the coupling constant. Hence, by using the repeated interaction protocol, we obtain the LME
\begin{eqnarray}
    \frac{d\rho}{dt} &=& -i[H_{s},\rho] + (\gamma_{L}^{-}+\gamma_{L}^{+})(\sigma_{1}^{x}\rho\sigma_{1}^{x}-\rho) \nonumber\\
    && + (\gamma_{R}^{-}+\gamma_{R}^{+})(\sigma_{2}^{x}\rho\sigma_{2}^{x}-\rho)\quad, \nonumber\\
    && \equiv -i[H_{s},\rho] + \mathcal{D}_{L}(\rho) + \mathcal{D}_{R}(\rho)~,
\end{eqnarray}
with $\gamma_{L(R)}^{-}=(1+n_{L(R)})g_{L(R)}^{2}$ and $\gamma_{L(R)}^{+}=n_{L(R)}g_{L(R)}^{2}$; $n_{L(R)}$ is the Bose-Einstein distribution for the $L(R)$ baths
$$
n_{L(R)} = \left[ e^{\omega_{L(R)}\beta_{L(R)}} - 1\right]^{-1}~.
$$

According to our previous definition, the heat and work currents becomes
\begin{equation*}
\begin{aligned}
    &\Dot{Q}_{L(R)}=\frac{1}{2}\left<[V_{L(R)},[V_{L(R)},H_{L(R)}]]\right>~,\\
    &\Dot{W}_{L(R)}=-\frac{1}{2}\left<[V_{L(R)},[V_{L(R)},H_{L(R)}+H_{s}]]\right>~.
\end{aligned}
\end{equation*}

Performing the computation, we obtain
\begin{equation*}
    \begin{aligned}
        &\Dot{Q}_{L(R)}= -g_{L(R)}^{2}\omega_{L(R)}~,\\
        &\Dot{W}_{L(R)}= g_{L(R)}^{2}\omega_{L(R)} -2g_{L(R)}^{2}h_{1(2)}^{'}(2n_{L(R)}+1)Tr^{S}(\sigma_{1(2)}^{z}\rho)\\
        &-2g_{L(R)}^{2}\Delta_{1,2}^{'}(2n_{L(R)}+1)Tr^{S}(\sigma_{1}^{z}\sigma_{2}^{z}\rho)\quad,
    \end{aligned}
\end{equation*}
where $\Delta' = \Delta/2, h'= h/2$.

We need to know the steady distribution to continue. We note that a diagonal matrix solution is possible: in this case, from the LME above, we have
\begin{equation*}
    \frac{d\rho_{jj}}{dt}=0=\mathcal{D}_{L}(\rho)_{jj}+\mathcal{D}_{R}(\rho)_{jj}~,
\end{equation*}
which gives us four equations
\begin{equation}
    \begin{aligned}
        &0=-(\alpha_{L}+\alpha_{R})\rho_{11}+\alpha_{R}\rho_{22}+\quad\alpha_{L}\rho_{33}+\quad 0\\
        &0=\quad\quad\alpha_{R}\rho_{11}-(\alpha_{L}+\alpha_{R})\rho_{22}+0\quad+\alpha_{L}\rho_{44}\\
        &0=\quad\quad\alpha_{L}\rho_{11}+\quad\quad0\quad\quad-(\alpha_{L}+\alpha_{R})\rho_{33}+\alpha_{R}\rho_{44}\\
        &0=\quad\quad 0\quad\quad+\alpha_{L}\rho_{22}\quad\quad+\alpha_{R}\rho_{33}\quad\quad-(\alpha_{L}+\alpha_{R})\rho_{44}\quad,\\
    \end{aligned}
\end{equation}
where $\alpha_{L(R)} = \gamma_{L(R)}^{-}+\gamma_{L(R)}^{+}$. These equations, together with the normalization $\sum_{i=1}^{4} \rho_{ii} = 1$ gives us $\rho_{11} = \rho_{22} = \rho_{33} =
\rho_{44} = 1/4$. Hence, we obtain
\begin{equation}
 \Dot{Q}_{L(R)} = -g_{L(R)}^{2}\omega_{L(R)}~, ~~~~\Dot{W}_{L(R)}=g_{L(R)}^{2}\omega_{L(R)}~.
\end{equation}
In resume, the total energy flow is zero ($\dot{Q}_{L} + \dot{W}_{L} =0$), but we have heat and work currents. We emphasize that we find, in the literature of boundary driven systems, results associated to heat instead of
associated to energy current \cite{EPL-Levy, Mendoza-A}; here we show that, even in simple systems, it may lead to incorrect statements.

We also carry out the computation for $N=3$. It is quite pertinent, since we may find sometimes considerable differences between $N=2$ (say, a junction) and $N>2$, see e.g. Ref.\cite{LOK}. Now, the Hamiltonian is
\begin{equation}
\begin{aligned} H_{s} =
&\frac{h_{1}}{2}\sigma_{1}^{z}+\frac{h_{2}}{2}\sigma_{2}^{z}+\frac{h_{3}}{2}\sigma_{3}^{z} \nonumber\\
&+ \frac{\Delta_{1,2}}{2}\sigma_{1}^{z}\sigma_{2}^{z} + \frac{\Delta_{2,3}}{2}\sigma_{2}^{z}\sigma_{3}^{z} +\frac{\Delta_{1,3}}{2}\sigma_{1}^{z}\sigma_{3}^{z}~,
\end{aligned}
\end{equation}
We introduce the ``long-range'' interaction $\Delta_{1,3}$ following a previous study on the quantum Ising model (however under weak interaction with the baths) in which a energy current is nonvanishing  only if the
first and the last sites are linked by an interaction, see Ref.\cite{EP2019}.

Carrying out the computation for the heat and work currents, we find
\begin{equation*}
        \Dot{Q}_{L(R)}=-g_{L(R)}^{2}\omega_{L(R)}~,
\end{equation*}
\begin{equation*}
\begin{aligned}
\Dot{W}_{L(R)}=&g_{L(R)}^{2}\omega_{L(R)}-2g_{L(R)}^{2}h_{1(3)}^{'}(2n_{L(R)}+1)Tr^{s}(\sigma_{1(3)}^{z}\rho\\
&-2g_{L(R)}^{2}\Delta_{1(2),2(3)}^{'}(2n_{L(R)}+1)Tr^{S}(\sigma_{1(2)}^{z}\sigma_{2(3)}^{z}\rho)\\
&-2g_{L}^{2}\Delta_{1,3}^{'}Tr^{S}\sigma_{1}^{z}\sigma_{3}^{z}\rho ~,
\end{aligned}
\end{equation*}
where $\Delta' = \Delta/2, h'=h/2$.
Again, we need to know the steady state distribution to go on. As before, we take a solution given by a diagonal matrix. We find two groups of equations, similar to the case $N=2$, one group for
$\rho_{1,1}, \rho_{2,2}, \rho_{5,5}, \rho_{6,6}$, and the other one for $\rho_{3,3}, \rho_{4,4}, \rho_{7,7}$ and $\rho_{8,8}$. Thus, from these equations and from $\sum_{i} \rho_{i,i} = 1$, we find
\begin{equation*}
    \begin{aligned}
            &\rho_{I}=\rho_{11}=\rho_{22}=\rho_{55}=\rho_{66}= \frac{a}{4}~,\\
            &\rho_{II}=\rho_{33}=\rho_{44}=\rho_{77}=\rho_{88}=\frac{b}{4}~,
    \end{aligned}
\end{equation*}
where $a$ and $b$ are arbitrary positive numbers with $a + b = 1$.

Turning to the expressions for heat and work currents, we get (again)
\begin{equation*}
 \Dot{Q}_{L(R)} = -g_{L(R)}^{2}\omega_{L(R)}~, ~~~~\Dot{W}_{L(R)}=g_{L(R)}^{2}\omega_{L(R)}~.
\end{equation*}
That is, there is no total energy flow, but there are heat and work currents.

Now we turn to the case of a recurrently used type of environment: the spin bath. We take, as baths, two extra spins, each one coupled to an end of the chain system. For the Hamiltonian of the baths we take $H_{L(R)}=h_{L(R)}\sigma_{L(R)}^{z}/2$, and for interaction between system and baths
\begin{equation*}
V_{L(R)}=\sqrt{\frac{\gamma_{L(R)}}{\tau}}\left(\sigma_{L(R)}^{x}\sigma_{1(N)}^{x}+\sigma_{1(N)}^{y}\sigma_{L(R)}^{y}\right).
\end{equation*}
By using the repeated interaction protocol, we get the LME
\begin{equation*}
\frac{d\rho}{dt}=-i[H_{s},\rho]+\sum_{r=L,R}\mathcal{D}_{r}(\rho)\quad,
\end{equation*}
\begin{equation*}
\begin{aligned}
\mathcal{D}_{L(R)}=&X_{L(R)}^{+}\left[\sigma_{1(N)}^{+}\rho\sigma_{1(N)}^{-}-\frac{1}{2}\{\sigma_{1(N)}^{-}\sigma_{1(N)}^{+},\rho\}\right]+\\
+&X_{L(R)}^{-}\left[\sigma_{1(N)}^{-}\rho\sigma_{1(N)}^{+}-\frac{1}{2}\{\sigma_{1(N)}^{+}\sigma_{1(N)}^{-},\rho\}\right]]
\end{aligned}
\end{equation*}
where $X_{L(R)}^{\pm}=2\gamma_{L(R)}(1\pm f_{L(R)})$; $\{\cdot,\cdot\}$ is the anticommutator; and
\begin{equation*}
f_{L(R)}=\left<\sigma_{L(R)}^{z}\right>=-tanh\left(\beta_{L(R)}\frac{h_{L(R)}}{2}\right)~.
\end{equation*}

For the chain with $N=2$, we obtain for the heat and work currents
\begin{equation}
\Dot{Q}_{L(R)}=2\gamma_{L(R)} h_{L(R)}\left[f_{L(R)}-Tr^{S}\sigma_{1(2)}^{z}\rho\right]~,
\end{equation}
\begin{equation}
\begin{aligned}
\Dot{W}_{L(R)}=&2\gamma_{L(R)} h_{1(2)}\left[f_{L(R)}-Tr^{S}\sigma_{1(2)}^{z}\rho\right] \\
&-2\gamma_{L(R)} h_{L(R)}\left[f_{L(R)}-Tr^{S}\sigma_{1(2)}^{z}\rho\right]\\
&-2\gamma_{L(R)}\Delta_{1,2}Tr^{S}\sigma_{1}^{z}\sigma_{2}^{z}\rho\\ &+2\gamma_{L(R)} \Delta_{1,2}f_{L(R)}Tr^{S}\sigma_{2(1)}^{z}\rho~.
\end{aligned}
\end{equation}

Again, we find a diagonal steady state density matrix by solving the LME
\begin{equation*}
\frac{d\rho_{jj}}{dt}=0=\mathcal{D}_{L}(\rho)_{jj}+\mathcal{D}_{R}(\rho)_{jj}~.
\end{equation*}
we obtain
\begin{equation*}
\begin{aligned}
&0=-(X_{L}^{+}+X_{R}^{-})\rho_{22}+X_{R}^{+}\rho_{11}+\quad 0+\quad X_{L}^{-}\rho_{44}\\
&0=\quad X_{R}^{-}\rho_{22}-(X_{L}^{+}+X_{R}^{+})\rho_{11}+X_{L}^{-}\rho_{33}+0\quad\\
&0=\quad 0+\quad X_{L}^{+}\rho_{11}-\quad(X_{L}^{-}+X_{R}^{+})\rho_{33}+X_{R}^{-}\rho_{44}\\
&0=\quad X_{L}^{+}\rho_{22}+\quad 0\quad+X_{R}^{+}\rho_{33}\quad-(X_{L}^{-}+X_{R}^{-})\rho_{44}\quad.\\
\end{aligned}
\end{equation*}
From these equations, together with $\sum_{i}\rho_{i,i} =1$, we find
\begin{equation*}
\rho=\begin{bmatrix}
\frac{X_{R}^{-}X_{L}^{-}}{16\gamma_{L}\gamma_{R}} & 0 & 0 & 0 \\
0&\frac{X_{R}^{+}X_{L}^{-}}{16\gamma_{L}\gamma_{R}}  &0  & 0\\
0& 0 &\frac{X_{R}^{-}X_{L}^{+}}{16\gamma_{L}\gamma_{R}}  &0 \\
0& 0 &0  &\frac{X_{R}^{+}X_{L}^{+}}{16\gamma_{L}\gamma_{R}}
\end{bmatrix}~.
\end{equation*}
Hence, with such density matrix, we obtain
\begin{equation*}
Tr_{s}\sigma_{1}^{z}\rho=f_{L}~, ~~
Tr_{s}\sigma_{1}^{z}\sigma_{2}^{z}\rho=f_{L}f_{R}~, ~~
Tr_{s}\sigma_{2}^{z}\rho=f_{R}~.
\end{equation*}
And finally, for the heat and work currents,
\begin{equation*}
\Dot{Q}_{L(R)}=0~,~~
\Dot{W}_{L(R)}=0~.
\end{equation*}
That is, now both currents also vanish.

We carry out the computation for $N=3$ and interaction between the first and
last sites, as in the previous case of bosonic baths. Again, after some algebra, we obtain $\Dot{Q}_{L(R)}=0$ and $\Dot{W}_{L(R)}=0$.

\section{Final Remarks}

Considering thermodynamics aspects, it is pertinent to comment on the entropy rate. We denote by $S_{S}(t) = -Tr_{S}\{\rho_{S}(t)\ln \rho_{S}(t)\}$ the von
Neumann entropy of the system. The system entropy production is given by the
difference between the change in the von Neumann entropy and the baths entropy flux \cite{FBarra, EspoPRX}
$$
\Pi = \frac{dS_{S}}{dt} - \sum_{r} \beta_{r}\dot{Q}_{r} ~.
$$
In the steady state, as the entropy of the system is constant, we have
$$
\Pi_{SS} = -\sum_{r} \beta_{r}\dot{Q}_{r} ~.
$$
Moreover, for these quantum spin chains, see Refs.\cite{FBarra, Pereira2018}, we can write it in terms of the spin current $J$
$$
\Pi_{SS} \propto (h_{R}\beta_{R} - h_{L}\beta_{L})J~.
$$
For the case of a $XXZ$ chain with three sites, the spin current $J$ is exactly computed in Ref.\cite{SPL}. For the case of $f = f_{L} = -\tanh(\beta_{L}h_{L}/2)$ and $f_{R}= -f= -\tanh(\beta_{R}h_{R}/2)$, we obtain $J$ proportional to $f$, for small values of $f$. Thus, in such a situation, we have $\beta_{R}h_{R} = -\beta_{L}h_{L}$, and $f \propto \beta_{R}h_{R}$. Consequently,
 $$
 \Pi_{SS} \propto (h_{R}\beta_{R} + h_{R}\beta_{R})h_{R}\beta_{R}
 \propto (h_{R}\beta_{R})^{2} \geq 0~.
  $$

For the steady state of the quantum Ising model, for bosonic baths and
spin-boson interaction, we obtain
$$
\Pi_{SS} = \beta_{L}g_{L}^{2}\omega_{L} + \beta_{R}g_{R}^{2}\omega_{R} \geq 0~.
$$

These results express the second law of thermodynamics.

%%%%%%%%%%%%%%%%%%%%%%%%%%%%%%%%%%%%%%%%%%%%%%%%%%%%%%%%%%%%%%%%%%%%%%%%%%%%%%%%%%%%%%%%%%%%%%%%%%%%%%%%%%%%%%%%%%%%%%%%%%%%%%%%%%%%%%%%%%%%%%%%%%%%%%%%%%%%%%%%%%%%

\appendix

\section{The RI protocol and the LME}

For clearness and completeness we repeat here some manipulations detailed described in previous papers \cite{FBarra, LOK}.

To obtain the LME from the RI protocol we turn to discrete mapping presented in eq.(\ref{map}). Denoting $\rho_{T} = \rho_{S}\rho_{E}$ and using the Baker-Campbell-Hausdorff formula
\begin{eqnarray*}
U\rho_{T}U^{\dagger} &=& e^{-iH_{T}\tau}\rho_{T}e^{iH_{T}\tau} = \\
 &=& \rho_{T} - i\tau[H_{T},\rho_{T}] - \frac{\tau^{2}}{2} [H_{T}, [H_{T},\rho_{T}]] + \ldots .
 \end{eqnarray*}
Then, we introduce it in eq.(\ref{map}) and take the partial trace over the baths. For the first term we obtain $Tr_{L,R}(\rho_{L}\rho_{S}((n-1)\tau)\rho_{R}) = \rho_{S}((n-1)\tau)$. For the second,
$Tr_{L,R}\{[H_{T}, \rho_{T}]\} = [H_{S}, \rho((n-1)\tau)]$. For the last, in order to keep a nonvanishing interaction in the limit $\tau\rightarrow 0$ to be considered ahead, we need to rescale $V$ to make it increasing with
$\tau$. We write, e.g., for $V_{L}$,
$$
V_{L} = \sqrt{\frac{\lambda_{L}}{\tau}} \left( \sigma_{L}^{x}\sigma_{1}^{x} + \sigma_{L}^{y}\sigma_{1}^{y} \right)~.
$$
Hence, defining
$$
-\frac{\tau^{2}}{2}Tr_{L} [ V_{L}, [V_{L}, \rho_{T}]] \equiv \mathcal{D}_{L} ~,
$$
and similarly for $\mathcal{D}_{R}$, after some algebra we obtain
\begin{eqnarray*}
\rho_{S}(n\tau) &=& \rho_{S}((n-1)\tau)
 -i\tau [H_{S}, \rho_{S}((n-1)\tau)] + \\
 && \tau\left[\mathcal{D}_{L}(\rho_{S}) + \mathcal{D}_{R}(\rho_{S})\right] + \mathcal{O}(\tau^{>1})~.
\end{eqnarray*}
Taking the difference $\rho_{S}(n\tau) - \rho_{S}((n-1)\tau)$, dividing by $\tau$ and taking the limit $\tau\rightarrow 0$, we get the LME.

\section{Steady state analysis}

We investigate two different systems in this paper: the quantum Ising and the $XXZ$ model. Interested in phenomena such as energy rectification, we also consider asymmetric versions of the models. For the quantum
Ising model, the analysis is simpler and a direct computation involving the LME allows us to get the steady density matrix. For the $XXZ$ case, the situation is more intricate, and we describe here the method to find its
steady density matrix.

First, we introduce the transformation $vec(A)$, that transforms the matrix $A$ into a column vector by stacking the columns of the matrix. Precisely,
\begin{eqnarray*}
A = \left
[\begin{array}{cc} a & b \\
 c & d \end{array} \right]
~~\Rightarrow vec(A) = \left ( \begin{array}{c} a \\
c \\ b \\ d \end{array} \right ) ~.
\end{eqnarray*}
Before analyzing the LME, we recall an important identity for the vectorization: for any three $N\times N$ matrices, we have
$$
vec(ABC) = (C^{T}\otimes A) vec(B) ~.
$$
The symbol $T$ above means the transposition. Now, for the density matrix $\rho$, let us define the vector of size $d^{2} = 2^{2N}$
$$
|\rho\rangle \equiv vec(\rho)~.
$$
Turning to the LME, the right-hand side of the equation may be written in terms of the product of three matrices $2^{N}\times 2^{N}$ such as $a\rho B$ if we take the terms $A\rho$ and $\rho B$ as $A\rho I$ and $I\rho B$, where $I$ is the identity. Hence, with the vectorization, the LME become the linear equation
$$
\frac{d|\rho\rangle}{dt} = M |\rho\rangle ~,
$$
where the linear operator $M$ is a matrix of size $2^{2N}\times 2^{2N}$. The steady state is the eigenstate of $M$ with eigenvalue zero. And, for the $XXZ$ model considered here, it is unique \cite{ProsenS}.

Then, following the method, we can, in principle, find exact results for the density matrix. However, as the size of $M$ rapidly increases with $N$, it is hard to obtain even numerical results for asymmetric matrices and large
$N$. Here, we will solve exactly only a small chain of size $N=3$ (we give also results and comments on $N=2$), which is already enough to establish our main message: we must go beyond the LME in order to obtain precise
information about the energy flow and related issues (heat and work).

\section{Some formulas for heat and work currents}

As illustration, we describe below some formulas found for heat and work currents in the case of Hamiltonian (\ref{Hamiltonian2}).

For $N=3$, $f_{L}= f = -f_{R}$, we have
\begin{widetext}
\begin{equation*}
\begin{aligned}
	\dot{Q}_L &= \big[2 \gamma  f \alpha^2 h_L ((\gamma ^2+\delta ^2)^2 (81 \gamma ^4-18 \gamma ^2 (2 f^2-3)
		(\delta ^2+\Delta ^2)+(3-2 f^2)^2 (\delta ^2-\Delta ^2)^2)+\alpha^6 (72 \gamma ^2
		\nonumber \\ & \quad +3 \Delta
		^2+4 \delta ^2 (5 f^2+9))+\alpha^2 (\gamma ^2+\delta ^2) (216 \gamma ^4-3 \gamma ^2 (4 \delta ^2
		(f^2-9)+\Delta ^2 (16 f^2-27))-(2 f^2
		\nonumber \\ & \quad -3) (\Delta ^4+4 \delta ^4 (f^2+1)+\delta ^2
		\Delta ^2 (13-4 f^2)))+2 \alpha^4 (99 \gamma ^4+\gamma ^2 (2 \delta ^2 (7 f^2+48)+3 \Delta
		^2 (5-2 f^2))
		\nonumber \\ & \quad -3 \delta ^2 \Delta ^2 (f^2-4)+\delta ^4 (2 f^4-6 f^2+15))+9
		\alpha^8)\big]\big[\alpha^8 (81 \gamma ^2+3 \Delta ^2+\delta ^2 (20 f^2+39))+(\gamma ^2
		\nonumber \\ & \quad +\delta ^2)^2
		(81 \gamma ^6+\gamma ^2 (\delta ^4 (27-8 f^4)+2 \delta ^2 \Delta ^2 (8 f^4+9)+\Delta ^4 (27-8
		f^4))+9 \gamma ^4 (2 f^2+9) (\delta ^2+\Delta ^2)
		\nonumber \\ & \quad -(2 f^2-3) (\delta ^2-\Delta
		^2)^2 (\delta ^2+\Delta ^2))+2 \alpha^6 (135 \gamma ^4+3 \gamma ^2 (7 \delta ^2
		(f^2+6)+\Delta ^2 (7-2 f^2))
		\nonumber \\ & \quad +\delta ^2 \Delta ^2 (17-3 f^2)+\delta ^4 (2 f^4-3
		f^2+21))+2 \alpha^4 (207 \gamma ^6+\gamma ^4 (\delta ^2 (291-7 f^2)+3 \Delta ^2 (26-7
		f^2))
		\nonumber \\ & \quad+\gamma ^2 (-\Delta ^4 (f^2-3)+\delta ^4 (-8 f^4-16 f^2+107)+\delta ^2 \Delta ^2 (4
		f^4-37 f^2+104))-\delta ^2 (\Delta ^4 (f^2-3)
		\nonumber \\ & \quad +\delta ^4 (4 f^4+f^2-11)-2 \delta ^2 \Delta ^2
		(2 f^4-9 f^2+14)))+\alpha^2 (\gamma ^2+\delta ^2) (297 \gamma ^6-3 \gamma ^4 (\delta ^2
		(22 f^2-105)
		\nonumber \\ & \quad +2 \Delta ^2 (2 f^2-33))+\gamma ^2 (\delta ^4 (20 f^4-44 f^2+111)-4 \delta ^2
		\Delta ^2 (6 f^4+2 f^2-33)
		+\Delta ^4 (4 f^4-20 f^2
		\nonumber \\ & \quad +33))+\delta ^6 (4 f^4-10 f^2+13)-2 \delta ^4
		\Delta ^2 (4 f^4-14 f^2+1)+\delta ^2 \Delta ^4 (4 f^4-18 f^2+25))+9 \alpha^{10}\big]^{-1}
\end{aligned}
\end{equation*}
\begin{equation*}
\begin{aligned}
	\dot{Q}_R &= -\big[2 \gamma  f \alpha^2 h_R ((\gamma ^2+\delta ^2)^2 (81 \gamma ^4-18 \gamma ^2 (2 f^2-3)
		(\delta ^2+\Delta ^2)+(3-2 f^2)^2 (\delta ^2-\Delta ^2)^2)+\alpha^6 (72 \gamma ^2
		\nonumber \\ & \quad +3 \Delta
		^2+4 \delta ^2 (5 f^2+9))+\alpha^2 (\gamma ^2+\delta ^2) (216 \gamma ^4-3 \gamma ^2 (4 \delta ^2
		(f^2-9)+\Delta ^2 (16 f^2-27))-(2 f^2
		\nonumber \\ & \quad -3) (\Delta ^4+4 \delta ^4 (f^2+1)+\delta ^2
		\Delta ^2 (13-4 f^2)))+2 \alpha^4 (99 \gamma ^4+\gamma ^2 (2 \delta ^2 (7 f^2+48)+3 \Delta
		^2 (5-2 f^2))
		\nonumber \\ & \quad -3 \delta ^2 \Delta ^2 (f^2-4)+\delta ^4 (2 f^4-6 f^2+15))+9
		\alpha^8)\big]\big[\alpha^8 (81 \gamma ^2+3 \Delta ^2+\delta ^2 (20 f^2+39))+(\gamma ^2
		\nonumber \\ & \quad +\delta ^2)^2
		(81 \gamma ^6+\gamma ^2 (\delta ^4 (27-8 f^4)+2 \delta ^2 \Delta ^2 (8 f^4+9)+\Delta ^4 (27-8
		f^4))+9 \gamma ^4 (2 f^2+9) (\delta ^2+\Delta ^2)
		\nonumber \\ & \quad -(2 f^2-3) (\delta ^2-\Delta
		^2)^2 (\delta ^2+\Delta ^2))+2 \alpha^6 (135 \gamma ^4+3 \gamma ^2 (7 \delta ^2
		(f^2+6)+\Delta ^2 (7-2 f^2))+\delta ^2 \Delta ^2 (17
		\nonumber \\ & \quad -3 f^2)+\delta ^4 (2 f^4-3
		f^2+21))+2 \alpha^4 (207 \gamma ^6+\gamma ^4 (\delta ^2 (291-7 f^2)+3 \Delta ^2 (26-7
		f^2))
		\nonumber \\ & \quad +\gamma ^2 (-\Delta ^4 (f^2-3)+\delta ^4 (-8 f^4-16 f^2+107)+\delta ^2 \Delta ^2 (4
		f^4-37 f^2+104))-\delta ^2 (\Delta ^4 (f^2-3)
		\nonumber \\ & \quad +\delta ^4 (4 f^4+f^2-11)-2 \delta ^2 \Delta ^2
		(2 f^4-9 f^2+14)))+\alpha^2 (\gamma ^2+\delta ^2) (297 \gamma ^6-3 \gamma ^4 (\delta ^2
		(22 f^2-105)
		\nonumber \\ & \quad +2 \Delta ^2 (2 f^2-33))+\gamma ^2 (\delta ^4 (20 f^4-44 f^2+111)-4 \delta ^2
		\Delta ^2 (6 f^4+2 f^2-33)+\Delta ^4 (4 f^4-20 f^2
		\nonumber \\ & \quad +33))+\delta ^6 (4 f^4-10 f^2+13)-2 \delta ^4
		\Delta ^2 (4 f^4-14 f^2+1)+\delta ^2 \Delta ^4 (4 f^4-18 f^2+25))+9 \alpha^{10}\big]^{-1}
\end{aligned}
\end{equation*}
\begin{equation*}
\begin{aligned}
	\dot{W}_L &= -\big[2 f \alpha^2 \gamma  (9 h_L \alpha^8+(h_L (72 \gamma ^2+4 (5 f^2+9) \delta ^2+3 \Delta
		^2)-12 f \delta  \Delta ^2) \alpha^6+2 (2 f \delta  (9 \gamma ^4+2 (5 \delta ^2
		\nonumber \\ & \quad -6 f^2 \Delta ^2)
		\gamma ^2+\delta ^4-\Delta ^4-2 (f^2-3) \delta ^2 \Delta ^2)+h_L (99 \gamma ^4+(2 (7 f^2+48)
		\delta ^2+3 (5-2 f^2) \Delta ^2) \gamma ^2
		\nonumber \\ & \quad +(2 f^4 -6 f^2+15) \delta ^4-3 (f^2-4) \delta ^2 \Delta
		^2)) \alpha^4+(\gamma ^2+\delta ^2) (4 f \delta  (54 \gamma ^4+3 (2 (3 f^2+7) \delta
		^2+(5
		\nonumber \\ & \quad -6 f^2) \Delta ^2) \gamma ^2
		 +2 (f^2+2) \delta ^4+(1-2 f^2) \Delta ^4+13 \delta ^2 \Delta
		^2)+h_L (216 \gamma ^4-3 (4 (f^2-9) \delta ^2+(16 f^2
		\nonumber \\ & \quad -27) \Delta ^2) \gamma ^2 -(2f^2-3) (4 (f^2+1) \delta ^4+(13-4 f^2) \Delta ^2 \delta ^2+\Delta ^4)))
		\alpha^2+(\gamma ^2+\delta ^2)^2 (h_L (81 \gamma ^4
		\nonumber \\ & \quad -18 (2 f^2-3) (\delta ^2 +\Delta ^2)
		\gamma ^2+(3-2 f^2)^2 (\delta ^2-\Delta ^2)^2)-4 f \delta  (-81 \gamma ^4+18 ((f^2-2)
		\delta ^2-f^2 \Delta ^2) \gamma ^2
		\nonumber \\ & \quad +(2 f^2-3) (\delta ^2-\Delta ^2)^2))-h (9
		\alpha^8+(72 \gamma ^2+4 (5 f^2+9) \delta ^2+3 \Delta ^2) \alpha^6+2 (99 \gamma ^4+(2 (7
		f^2
		\nonumber \\ & \quad +48) \delta ^2+3 (5-2 f^2) \Delta ^2) \gamma ^2+(2 f^4-6 f^2+15) \delta ^4-3 (f^2-4)
		\delta ^2 \Delta ^2) \alpha^4+(\gamma ^2+\delta ^2) (216 \gamma ^4
		\nonumber \\ & \quad -3 (4 (f^2-9) \delta
		^2+(16 f^2-27) \Delta ^2) \gamma ^2-(2 f^2-3) (4 (f^2+1) \delta ^4+(13-4 f^2)
		\Delta ^2 \delta ^2+\Delta ^4)) \alpha^2
		\nonumber \\ & \quad +(\gamma ^2+\delta ^2)^2 (81 \gamma ^4-18 (2 f^2-3)
		(\delta ^2+\Delta ^2) \gamma ^2+(3-2 f^2)^2 (\delta ^2-\Delta ^2)^2)))\big]\big[9
		\alpha^{10}+(81 \gamma ^2
		\nonumber \\ & \quad +(20 f^2+39) \delta ^2+3 \Delta ^2) \alpha^8+2 (135 \gamma ^4+3 (7
		(f^2+6) \delta ^2+(7-2 f^2) \Delta ^2) \gamma ^2+(2 f^4-3 f^2
		\nonumber \\ & \quad +21) \delta ^4+(17-3 f^2)
		\delta ^2 \Delta ^2) \alpha^6+2 (207 \gamma ^6+((291-7 f^2) \delta ^2+3 (26-7 f^2) \Delta
		^2) \gamma ^4+((-8 f^4
		\nonumber \\ & \quad -16 f^2+107) \delta ^4+(4 f^4-37 f^2+104) \Delta ^2 \delta ^2-(f^2-3)
		\Delta ^4) \gamma ^2-\delta ^2 ((4 f^4+f^2-11) \delta ^4
		\nonumber \\ & \quad -2 (2 f^4-9 f^2+14) \Delta ^2 \delta
		^2+(f^2-3) \Delta ^4)) \alpha^4+(\gamma ^2+\delta ^2) (297 \gamma ^6-3 ((22
		f^2-105) \delta ^2
		\nonumber \\ & \quad +2 (2 f^2-33) \Delta ^2) \gamma ^4+((20 f^4-44 f^2+111) \delta ^4-4 (6 f^4+2
		f^2-33) \Delta ^2 \delta ^2+(4 f^4-20 f^2
		\nonumber \\ & \quad +33) \Delta ^4) \gamma ^2+(4 f^4-10 f^2+13) \delta ^6+(4
		f^4-18 f^2+25) \delta ^2 \Delta ^4-2 (4 f^4-14 f^2+1) \delta ^4 \Delta ^2) \alpha^2
		\nonumber \\ & \quad +(\gamma ^2+\delta
		^2)^2 (81 \gamma ^6+9 (2 f^2+9) (\delta ^2+\Delta ^2) \gamma ^4+((27-8 f^4) \delta ^4+2
		(8 f^4+9) \Delta ^2 \delta ^2+(27
		\nonumber \\ & \quad -8 f^4) \Delta ^4) \gamma ^2-(2 f^2-3) (\delta ^2-\Delta
		^2)^2 (\delta ^2+\Delta ^2))\big]^{-1}
\end{aligned}
\end{equation*}
\begin{equation*}
\begin{aligned}
	\dot{W}_R &= \big[2 f \alpha^2 \gamma  (9 h_R \alpha^8+(h_R (72 \gamma ^2+4 (5 f^2+9) \delta ^2+3 \Delta
		^2)-12 f \delta  \Delta ^2) \alpha^6+2 (2 f \delta  (9 \gamma ^4+2 (5 \delta ^2
		\nonumber \\ & \quad -6 f^2 \Delta ^2)
		\gamma ^2+\delta ^4-\Delta ^4-2 (f^2-3) \delta ^2 \Delta ^2)+h_R (99 \gamma ^4+(2 (7 f^2+48)
		\delta ^2+3 (5-2 f^2) \Delta ^2) \gamma ^2
		\nonumber \\ & \quad +(2 f^4-6 f^2+15) \delta ^4-3 (f^2-4) \delta ^2 \Delta
		^2)) \alpha^4+(\gamma ^2+\delta ^2) (4 f \delta  (54 \gamma ^4+3 (2 (3 f^2+7) \delta
		^2
		\nonumber \\ & \quad +(5-6 f^2) \Delta ^2) \gamma ^2+2 (f^2+2) \delta ^4+(1-2 f^2) \Delta ^4+13 \delta ^2 \Delta
		^2)+h_R (216 \gamma ^4-3 (4 (f^2-9) \delta ^2
		\nonumber \\ & \quad +(16 f^2-27) \Delta ^2) \gamma ^2-(2
		f^2-3) (4 (f^2+1) \delta ^4+(13-4 f^2) \Delta ^2 \delta ^2+\Delta ^4)))
		\alpha^2+(\gamma ^2
		\nonumber \\ & \quad +\delta ^2)^2 (h_R (81 \gamma ^4-18 (2 f^2-3) (\delta ^2+\Delta ^2)
		\gamma ^2+(3-2 f^2)^2 (\delta ^2-\Delta ^2)^2)-4 f \delta  (-81 \gamma ^4
		\nonumber \\ & \quad +18 ((f^2-2)
		\delta ^2-f^2 \Delta ^2) \gamma ^2+(2 f^2-3) (\delta ^2-\Delta ^2)^2))-h (9
		\alpha^8+(72 \gamma ^2+4 (5 f^2+9) \delta ^2
		\nonumber \\ & \quad +3 \Delta ^2) \alpha^6+2 (99 \gamma ^4+(2 (7
		f^2+48) \delta ^2+3 (5-2 f^2) \Delta ^2) \gamma ^2+(2 f^4-6 f^2+15) \delta ^4-3 (f^2
		\nonumber \\ & \quad -4)
		\delta ^2 \Delta ^2) \alpha^4+(\gamma ^2+\delta ^2) (216 \gamma ^4-3 (4 (f^2-9) \delta
		^2+(16 f^2-27) \Delta ^2) \gamma ^2-(2 f^2-3) (4 (f^2
		\nonumber \\ & \quad +1) \delta ^4+(13-4 f^2)
		\Delta ^2 \delta ^2+\Delta ^4)) \alpha^2+(\gamma ^2+\delta ^2)^2 (81 \gamma ^4-18 (2 f^2-3)
		(\delta ^2+\Delta ^2) \gamma ^2+(3
		\nonumber \\ & \quad -2 f^2)^2 (\delta ^2-\Delta ^2)^2)))\big]\big[9
		\alpha^{10}+(81 \gamma ^2+(20 f^2+39) \delta ^2+3 \Delta ^2) \alpha^8+2 (135 \gamma ^4+3 (7
		(f^2+6) \delta ^2
		\nonumber \\ & \quad +(7-2 f^2) \Delta ^2) \gamma ^2+(2 f^4-3 f^2+21) \delta ^4+(17-3 f^2)
		\delta ^2 \Delta ^2) \alpha^6+2 (207 \gamma ^6+((291-7 f^2) \delta ^2
		\nonumber \\ & \quad +3 (26-7 f^2) \Delta
		^2) \gamma ^4+((-8 f^4-16 f^2+107) \delta ^4+(4 f^4-37 f^2+104) \Delta ^2 \delta ^2-(f^2-3)
		\Delta ^4) \gamma ^2
		\nonumber \\ & \quad -\delta ^2 ((4 f^4+f^2-11) \delta ^4-2 (2 f^4-9 f^2+14) \Delta ^2 \delta
		^2+(f^2-3) \Delta ^4)) \alpha^4+(\gamma ^2+\delta ^2) (297 \gamma ^6
		\nonumber \\ & \quad -3 ((22
		f^2-105) \delta ^2+2 (2 f^2-33) \Delta ^2) \gamma ^4+((20 f^4-44 f^2+111) \delta ^4-4 (6 f^4+2
		f^2-33) \Delta ^2 \delta ^2
		\nonumber \\ & \quad +(4 f^4-20 f^2+33) \Delta ^4) \gamma ^2+(4 f^4-10 f^2+13) \delta ^6+(4
		f^4-18 f^2+25) \delta ^2 \Delta ^4-2 (4 f^4-14 f^2
		\nonumber \\ & \quad +1) \delta ^4 \Delta ^2) \alpha^2+(\gamma ^2+\delta
		^2)^2 (81 \gamma ^6+9 (2 f^2+9) (\delta ^2+\Delta ^2) \gamma ^4+((27-8 f^4) \delta ^4+2
		(8 f^4+9) \Delta ^2 \delta ^2
		\nonumber \\ & \quad +(27-8 f^4) \Delta ^4) \gamma ^2-(2 f^2-3) (\delta ^2-\Delta
		^2)^2 (\delta ^2+\Delta ^2))\big]^{-1}	
\end{aligned}
\end{equation*}

\end{widetext}

 \newpage

 %%%%%%%%%%%%%%%%%%%%%%%%%%%%%%%%%%%%%%%%%%%%%%%%%%%%%%%%%%%%%%%%%%%%%%%%%%%%%%%%%%%%%%%%%%%%%%%%%%%%%%%%%%%%%%%%%%%%%%%%%%%%%%%%%%%%%%%%%%%%%%%%%%%%%%%%%%%%%%%%%%%%%%%%%%%%%%%%%%%%%%%%%%%%%%%%%%%%%%%%%%%%%%%%%%%%%%%%
 %%%%%%%%%%%%%%%%%%%%%%%%%%%%%%%%%%%%%%%%%%%%%%%%%%%%%%%%%%%%%%%%%%%%%%%%%%%%%%%%%%%%%%%%%%%%%%%%%%%%%%%%%%%%%%%%%%%%%%%%%%%%%%%%%%%%%%%%%%%%%%%%%%%%%%%%%%%%%%%%%%%%%%%%%%%%%%%%%%%%%%%%%%%%%%%%%%%%%%%%%%%%%%%%%%%%%%%%%%%%

%%%%%%%%%%%%%%%%%%%%%%%%%%%%%%%%%%%%%%%%%%%%%%%%%%%%%%%%%%%%%%%%%%%%%%%%%%%%%%%%%%%%%%%%%%%%%%%%%%%%%%%%%%%%%%%%%%%%%%%%%%%%%%%%%%%%%%%%%%%%%%%%%%%%%%%%%%%%%%%%%%%%%%%%%%%%%%%%%%%%%%%%%%%%%%

%%%%%%%%%%%%%%%%%%%%%%%%%%%%%%%%%%%%%%%%%%%%%%%%%%%%%%%%%%%%%%%%%%%%%%%%%%%%%%%%%%%%%%%%%%%%%%%%%%%%%%%%%%%%%%%%%%%%%%%%%%%%%%%%%%%%%%%%%%%%%%%%%%%%%%%%%%%%%%%%%%%%%%%%%%%%%%%%%%%%%%%%%%%%%%

%%%%%%%%%%%%%%%%%%%%%%%%%%%%%%%%%%%%%%%%%%%%%%%%%%%%%%%%%%%%%%%%%%%%%%%%%%%%%%%%%%%%%%%%%%%%%%%%%%%%%%%%%%%%%%%%%%%%%%%%%%%%%%%%%%%%%%%%%%%%%%%%%%%%%%%%%%%%%%%%%%%%%%%%%%%%%%%%%%%%%%%%%%%%%%%%%

%%%%%%%%%%%%%%%%%%%%%%%%%%%%%%%%%%%%%%%%%%%%%%%%%%%%%%%%%%%%%%%%%%%%%%%%%%%%%%%%%%%%%%%%%%%%%%%%%%%%%%%%%%%%%%%%%%%%%%%%%%%%%%%%%%%%%%%%%%%%%%%%%%%%%%%%%%%%%%%%%%%%%%%%%%%%%%%%%%%%%%%%%%%%%%%%

%%%%%%%%%%%%%%%%%%%%%%%%%%%%%%%%%%%%%%%%%%%%%%%%%%%%%%%%%%%%%%%%%%%%%%%%%%%%%%%%%%%%%%%%%%%%%%%%%%%%%%%%%%%%%%%%%%%%%%%%%%%%%%%%%%%%%%%%%%%%%%%%%%%%%%%%%%%%%%%%%%%%%%

 {\bf Acknowledgments:} This work was partially supported by CNPq (Brazil).

\end{document}